\documentclass[aps,pra,twocolumn,a4paper,english]{revtex4}
\usepackage{amsmath,epsfig,amssymb,subfigure,bm,dsfont}
\usepackage{lipsum}

\begin{document}

\title{All-optical control of thermal conduction in waveguide QED}
\author{Wei-Bin Yan$^{1}$, Zhong-Xiao Man$^{1,\footnote{manzhongxiao@163.com}}$, Ying-Jie Zhang$%
^{1,\footnote{yingjiezhang@qfnu.edu.cn}}$, Heng Fan$^{2,\footnote{hfan@iphy.ac.cn}}$, and Yun-Jie Xia$^1 $}
\affiliation{$^1$College of Physics and Engineering, Qufu Normal University, Qufu,
273165, China}
\affiliation{$^2$Beijing National Laboratory for Condensed Matter Physics, Institute of
Physics, Chinese Academy of Sciences, Beijing 100190, China}

\begin{abstract}
We investigate the heat conduction between two one-dimension waveguides
intermediated by a Laser-driving atom. The Laser provides the optical
control on the heat conduction. The tunable asymmetric conduction of the
heat against the temperature gradient is realized. Assisted by the modulated
Laser, the heat conduction from either waveguide to the other waveguide can
be suppressed. Meanwhile, the conduction towards the direction opposite to
the suppressed one is gained. The heat currents can be significantly
amplified by the energy flow of the Laser. Moveover, the scheme can act like
a heat engine.
\end{abstract}

\maketitle

\section{Introduction}

The scheme composed by one-dimension (1D) waveguides coupled to emitters,
known as waveguide QED (Quantum Electrodynamics), has been studied
extensively \cite%
{Colloquium,fan07,shi,fanthero,zhengqed,longo,faninout,roy,zhou1,pla1,pla2,pla3,crys,Nanofibe,dark,ChargeNoise,Chaos,Ratchet,TopologyNonreciprocal, giantLambda,dipoleapproximation,ultrastrong,BoundPhotonicPairs,twogiant,Cavitylike,Boundstates,multiplecoupling,gaintn,Schrodinger,Selforder, Nonperturbative,Bellstates,nonMarkoviandynamics,Modeling,Multiphoton,Amplification,Transparency,Implications,MultipleFano,Entanglementpreserving,Phasemodulated,interference,Transmission,Photonscattering}%
. The 1D continuum guided photonic modes can be realized in several
nanoscale schemes, such as surface plasmon of metallic nanowires \cite%
{pla1,pla2,pla3}, photonic crystal waveguides \cite{crys}, optical
nanofibers \cite{Nanofibe}, and so on. Waveguide QED provides a potential
platform for the management of few photons through light-emitter interaction.

On the other hand, the management and reuse of the wasted heat is a
significant issue because the energy resources are limited and the energy
utilization in industry is always accompanied by the heat lost. Recently,
due to the development in engineering individual small scale devices, the
management of the thermal conduction in the small scale devices has
attracted extensive attention \cite%
{retc,optimal,Quantumthermaltransistor,energybackflow,threereservoirs,interactingspinlike, DynamicalCoulombblockade,Allthermaltransistor,ThermalNoise,qubitqutritcoupling,nonequilibriumtwoqubit, Dynamicallyinduced,nonequilibriumVtype,twoatom,wu,collisionmodels,utilizingthreequbits, MobilityEdges,Isingmodel,segmentedXXZchains,Searching,criticalheat,heattransport,effectof,heatflowres,transistorinsuperconducting, quantifyingheat,Opticallycontrolledquantum,bathspectralfiltering,nonthermalbath,incoupledcavities,Quantumthermaltransistors, ThermalTransistorEffec,Redfield approach,rectificationthroughnonlinear,Darlingtonair,DzyaloshinskiiMoriya,Ballisticthermal,Decoupledheatcharge,Steadyentanglement,Coherencehanced, diodeandcirculator,Entanglementenhanced,multitransistor,Geometrybasedcirculation,CycleFluxRanking,coupledquantumdots,threesuperconductingcircuit,Correlationenabled}%
. A large number of the works focus on the realization of the thermal
quantum devices analogous to the electronic components, such as the
diode-like \cite{retc,optimal} and transistor-like \cite%
{Quantumthermaltransistor} operations. In the thermal diode-like operations,
the heat conduction towards the fixed direction is significantly suppressed
while the opposite direction is allowed. It will be of interest to realize
the tunable thermal diode-like operation, in which the suppressed conduction
direction can be artificially selected by tunable parameters. More
interestingly, due to the experimental operability, waveguide QED may
provide a potential platform for the thermal devices, which needs to be
explored.

Though the individual quantum thermal device has been studied extensively
with great progress, it deserves a deep study how to integrate them for the
applicable purpose. Waveguide QED may provide a potential platform for this
based on two facts. One is that waveguide QED provides a potential candidate
for the quantum network \cite{qn} by integrating the waveguides and emitters
into a large system \cite{BoundPhotonicPairs}, in which the waveguides act
as the channels and the emitters as the nods. The other is that the
locations and numbers of the emitters coupled to the waveguide can be
accurately manipulated in most of the practical implementations. The latter
implies that waveguide QED would show significant advantages of individually
managing the thermal conduction of each quantum thermal device in the
integrated system. It is necessary to note that the individual management is
essential but challenging. A potential approach to overcome this is to
develop the control on the thermal conduction by optical modulation \cite%
{Opticallycontrolledquantum}, compared to the thermal modulation \cite%
{Quantumthermaltransistor}. Therefore, provided that the heat conduction in
waveguides can be managed by the optical signal acting on the emitter, it is
reasonable to believe that waveguide QED will provide significant help to
the integrated thermal quantum devices.

For these purposes, we propose to control the heat conduction by the optical
signals based on waveguide QED. In our scheme, the heat conduction between
two 1D waveguides is assisted by an intermediate emitter, which is coupled
to the continuum photonic modes of the two waveguides. We employ two weak
Laser beams to drive the intermediate emitter and show that the heat
conduction can be controlled by the Laser beams. The heat conduction shows
asymmetry against the temperature gradient for appropriate laser parameters.
In the ideal case, assisted by the optically modulated Laser, either
direction of the heat conduction between the two waveguides can be
prohibited while the opposite direction is gained. In the nonideal case,
either direction can be significantly suppressed while the opposite
direction is gained. This provides the rectification and switch of the heat
conduction by the optical modulation. The heat currents of the waveguides
largely change when the energy flow of the Laser changes slightly.
Therefore, the scheme shows the thermal transistor-like operation. Besides,
the scheme can convert the heat to work, which can be considered as the
engine-like operation. Our investigation provides a potential advantageous
quantum-optic platform to the thermal management.

This paper is organized as follows. In Sec. \ref{modelandHamiltonian}, we
introduce the scheme and Hamiltonian of the system. In Sec. \ref{QME}, we
obtain the time evolution of reduced density operator for the emitter in the
weak coupling regime and discuss the elements of the density matrix. In Sec. %
\ref{HeatCurrents}, we investigate the energy properties in the steady-state
regime. The control of heat conduction by optical modulation and the
engine-like operation are demonstrated. In Sec. \ref{conclusions}, the
conclusion and necessary discussions are made.

\section{Model and Hamiltonian}

\label{modelandHamiltonian}

\begin{figure}[t]
\includegraphics[width=8cm, height=5cm]{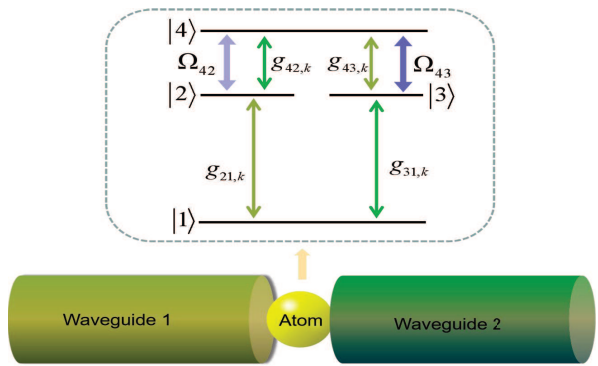}
\caption{An four-level atom driven by two Laser beams is side-coupled to two
1D waveguides.}
\label{model}
\end{figure}
The schematic diagram of the considered system is shown in Fig. \ref{model}.
A four-level emitter is side-coupled to two 1D waveguides. The emitter could
be a real atom or a manual atom-like object, such as the quantum dot \cite%
{crys} coupled to the plasmonic or photonic-crystal waveguide. The four
atomic states are denoted by $\left\vert j\right\rangle $ ($j=1,2,3,4$),
with the corresponding level frequencies $\omega _{j}$. The atomic
transitions $\left\vert 2\right\rangle \leftrightarrow \left\vert
1\right\rangle $ and $\left\vert 4\right\rangle \leftrightarrow \left\vert
3\right\rangle $ are coupled to the guided photons in the waveguide 1 with
strengths $g_{21,k}$ and $g_{43,k}$, respectively. The atomic transitions $%
\left\vert 3\right\rangle \leftrightarrow \left\vert 1\right\rangle $ and $%
\left\vert 4\right\rangle \leftrightarrow \left\vert 2\right\rangle $ are
coupled to the guided photons in the waveguide 2 with strengths $g_{31,k}$
and $g_{42,k}$, respectively. The symbol $k$ denotes the wave vector of the
guided photon. Here, we assume that the values of elements in the set $%
\{\omega _{43},\omega _{42}\}$ are obviously different from the ones in $%
\{\omega _{21},\omega _{31}\}$, with $\omega _{jj^{\prime }}=\omega
_{j}-\omega _{j^{\prime }}$ the atomic transition frequency. The external
laser beam with frequency $\omega _{L_{43}}$ ($\omega _{L_{42}}$) and Rabi
frequency $\Omega _{43}$ ($\Omega _{42}$) is introduced to drive the atomic
transition $\left\vert 4\right\rangle \leftrightarrow \left\vert
3\right\rangle $ ($\left\vert 4\right\rangle \leftrightarrow \left\vert
2\right\rangle $) through polarization and frequency selections. We assume
that the atom-waveguide and atom-Laser coupling strengths are much smaller
than the corresponding atomic transition frequencies. Hereafter, we take
reduced Plank constant $\hbar =1$. Within the rotating-wave approximation,
the Hamiltonian governing the system is%
\begin{equation*}
H=H_{A}+H_{W}+H_{A-W}+H_{A-L}
\end{equation*}%
with%
\begin{eqnarray}
H_{A} &=&\sum_{j=1...4}\omega _{j}\sigma ^{jj},  \notag \\
H_{W} &=&\sum_{k}\omega _{1,k}b_{1,k}^{\dagger }b_{1,k}+\sum_{k}\omega
_{2,k}b_{2,k}^{\dagger }b_{2,k},  \notag \\
H_{A-W} &=&\sum_{k}(g_{21,k}\sigma ^{21}b_{1,k}+g_{31,k}\sigma ^{31}b_{2,k}
\notag \\
&&+g_{43,k}\sigma ^{43}b_{1,k}+g_{42,k}\sigma ^{42}b_{2,k})+h.c.  \notag \\
H_{A-L} &=&\Omega _{42}\sigma ^{42}e^{-i\omega _{L42}t}+\Omega _{43}\sigma
^{43}e^{-i\omega _{L43}t}+h.c.  \label{Ham_shr}
\end{eqnarray}%
with the operator $\sigma ^{jj^{\prime }}=\left\vert j\right\rangle
\left\langle j^{\prime }\right\vert $ denoting the atomic raising, lowering,
and energy-level population operators, and the operator $b_{m,k}^{\dagger }$
($b_{m,k}$) creating (annihilating) a photon with wave vector $k$ in the
waveguide $m$. The symbol $\omega _{m,k}$ ($m=1,2$) denotes the photonic
frequency in waveguide $m$. We consider that the guided photonic frequency
is far from the waveguide cutoff frequency and hence the guided photonic
dispersion relation is approximately linearized. The Hamiltonian $H_{A}$
denotes the atomic energy, $H_{W}$ represents the energy of the guided
photon in the two waveguides, $H_{A-W}$ describes the coupling of the atom
to the two waveguides, and $H_{A-L}$ denotes the interaction between the
atom and the external Laser beams.

\section{Quantum Master Equation}

\label{QME}

The time evolution of the reduced density operator for the atom $\rho _{A}$
is obtained based on the approach represented in \cite{qo}. We assume that
the guided modes are distributed in the uncorrelated thermal equilibrium
mixture of states and hence the two waveguides are considered as two thermal
reservoirs. Consequently, one can obtain $\left\langle b_{m,k}\right\rangle
=\left\langle b_{m,k}^{\dagger }\right\rangle =0$, $\left\langle
b_{m,k}b_{m^{\prime },k^{\prime }}\right\rangle =\left\langle
b_{m,k}^{\dagger }b_{m^{\prime },k^{\prime }}^{\dagger }\right\rangle =0$, $%
\left\langle b_{m,k}b_{m^{\prime },k^{\prime }}^{\dagger }\right\rangle
=(n_{\omega _{k},m}+1)\delta _{m,m^{\prime }}\delta _{k,k^{\prime }}$, and $%
\left\langle b_{m,k}^{\dagger }b_{m^{\prime },k^{\prime }}\right\rangle
=n_{\omega _{k},m}\delta _{m,m^{\prime }}\delta _{k,k^{\prime }}$, with%
\begin{equation*}
n_{\omega _{k},m}=[\exp (\frac{\omega _{m,k}}{k_{B}T_{m}})-1]^{-1}
\end{equation*}%
the average occupation number of the photons with wave vector $k$ in the
reservoir $m$. The symbol $T_{m}$ represents temperature of the reservoir $m$
and $k_{B}$ denotes Boltzmann constant. The Hamiltonian (\ref{Ham_shr}) can
be mapped to the interaction picture as $%
H^{I}(t)=e^{iH_{0}t}H_{I}e^{-iH_{0}t}$, with $H_{0}=H_{A}+H_{W}$, and $%
H_{I}=H_{A-W}+H_{A-L}$. From the interaction picture von Neumann equation,
one can obtain%
\begin{eqnarray}
\dot{\rho}_{A}^{I}(t) &=&-iTr_{W}[H^{I}(t),\rho _{total}^{I}(t_{0})]  \notag
\\
&&-Tr_{W}\int_{t_{0}}^{t}dt^{\prime }[H^{I}(t),[H^{I}(t^{\prime }),\rho
_{total}^{I}(t^{\prime })]],  \label{von}
\end{eqnarray}%
where $\rho _{A}^{I}(t)$ represents the interaction-picture reduced density
operator for the atom, $\rho _{total}^{I}(t)$ is the interaction-picture
density operator of the atom-waveguide system, and $Tr_{W}$ denotes the
partial trace through the waveguides. Inserting $H^{I}(t)$ into Eqn. (\ref%
{von}) and performing Born, secular, and Weisskopf-Wigner approximations, we
can derive the master equation representing the time evolution of $\rho
_{A}^{I}(t)$ in the weak coupling regime (see Appendix for details). Turning
back to Schr\"{o}dinger picture, the evolution of the reduced matrix for the
atom is given as%
\begin{equation}
\dot{\rho}_{A}=-i[H_{A}+H_{A-L},\rho _{A}]+\sum_{lp}\mathcal{L}_{lp}[\rho
_{A}],  \label{MES}
\end{equation}%
with $lp\in \{21,31,42,43\}$. The dissipative Lindblad superoperator $%
\mathcal{L}_{lh}[\rho _{A}]$ has the form of%
\begin{eqnarray}
\mathcal{L}_{lp}[\rho _{A}] &=&\gamma _{lp}(n_{\omega _{lp}}+1)(\sigma
^{pl}\rho _{A}\sigma ^{lp}-\frac{1}{2}\{\sigma ^{ll},\rho _{A}\})  \notag \\
&&+\gamma _{lp}n_{\omega _{lp}}(\sigma ^{lp}\rho _{A}\sigma ^{pl}-\frac{1}{2}%
\{\sigma ^{pp},\rho _{A}\}),  \notag
\end{eqnarray}%
where $\gamma _{lp}$ denotes the atomic decay to the waveguide accompanied
by the transition $\left\vert l\right\rangle \leftrightarrow \left\vert
p\right\rangle $ (see Appendix for details). Here, we have labeled $%
n_{\omega _{jj^{\prime }},m}\rightarrow n_{\omega _{jj^{\prime }}}$ because
each atomic transition is coupled to its designated reservoir.

We focus on the steady-state regime. The master equation of Eqn. (\ref{MES})
contains the time-dependent phase. It will be convenient to look for a
rotating frame in which the steady-state reduced density operator is
time-independent. We bring in the rotating frame with respect to $H^{\prime
}=-\omega _{L42}\sigma ^{22}-\omega _{L43}\sigma ^{33}$. The arbitrary\
operator $O$ has the form of $O^{\prime }=e^{iH^{\prime }t}Oe^{-iH^{\prime
}t}$ in the rotating frame. In this frame, the atomic Hamiltonian $H_{A}$
keeps unchanged and the time evolution of the reduced density operator for
the atom turns out to be%
\begin{equation}
\dot{\rho}=-i[H_{A}-H^{\prime }+H_{A-L}^{\prime },\rho ]+\sum_{lp}\mathcal{L}%
_{lp}[\rho ],  \label{ME}
\end{equation}%
with $H_{A-L}^{\prime }=\Omega _{42}\sigma ^{42}+\Omega _{43}\sigma
^{43}+h.c.$ The Hamiltonian in the first term of the right hand in Eqn. (\ref%
{ME}) is time-independent. The time evolution of the density matrix diagonal
elements are obtained as%
\begin{eqnarray}
\dot{\rho}_{11} &=&\Gamma _{21}+\Gamma _{31},  \notag \\
\dot{\rho}_{22} &=&\Gamma _{42}-\Gamma _{21}-\Upsilon _{42},  \notag \\
\dot{\rho}_{33} &=&\Gamma _{43}-\Gamma _{31}-\Upsilon _{43},  \notag \\
\dot{\rho}_{44} &=&-\Gamma _{43}-\Gamma _{42}+\Upsilon _{42}+\Upsilon _{43},
\label{ele1}
\end{eqnarray}%
where $\rho _{jj^{\prime }}=\left\langle j\right\vert \rho \left\vert
j^{\prime }\right\rangle $, $\Gamma _{jj^{\prime }}={\gamma _{jj^{\prime }}[}%
\left( {n_{\omega _{jj^{\prime }}}}+1\right) \rho _{jj}-{n_{\omega
_{jj^{\prime }}}}\rho _{j^{\prime }j^{\prime }}]$ denotes the net decay rate
from $\left\vert j\right\rangle $ to $\left\vert j^{\prime }\right\rangle $
due to the atom-reservoir interaction, and $\Upsilon _{jj^{\prime }}=i\Omega
_{jj^{\prime }}\left( \rho _{jj^{\prime }}-\rho _{j^{\prime }j}\right) $ is
the net transition resulting from the atom-Laser interaction. The
off-diagonal elements $\rho _{42}$ and $\rho _{43}$ have the direct impact
on the time evolution of the diagonal elements, and $\rho _{32}$ has the
indirect impact on the diagonal elements. Their evolutions are obtained as%
\begin{eqnarray}
\dot{\rho}_{42} &=&-i(\omega _{42}-\omega _{L42})\rho _{42}+\alpha
_{42}-\eta _{432}  \notag \\
&&-\frac{\beta _{42}^{+}+\beta _{42}^{-}+\beta _{43}^{+}+\beta _{21}^{+}}{2}%
\rho _{42},  \notag \\
\dot{\rho}_{43} &=&-i(\omega _{43}-\omega _{L43})\rho _{43}+\alpha
_{43}-\eta _{423}  \notag \\
&&-\frac{\beta _{43}^{+}+\beta _{43}^{-}+\beta _{42}^{+}+\beta _{31}^{+}}{2}%
\rho _{43},  \notag \\
\dot{\rho _{32}} &=&-i[\omega _{32}-(\omega _{{L42}}-\omega _{{L43}})]\rho
_{32}-\eta _{243}^{\ast }+\eta _{342}^{\ast }  \notag \\
&&-\frac{\beta _{31}^{+}+\beta _{21}^{+}+\beta _{43}^{-}+\beta _{42}^{-}}{2}%
\rho _{32},  \label{ele2}
\end{eqnarray}%
with $\alpha _{jj^{\prime }}=i\Omega _{jj^{\prime }}\left( \rho _{jj}-\rho
_{j^{\prime }j^{\prime }}\right) $, $\beta _{jj^{\prime }}^{+}=({n_{\omega
_{jj^{\prime }}}}+1){\gamma _{jj^{\prime }}}$, $\beta _{jj^{\prime }}^{-}={%
n_{\omega _{jj^{\prime }}}\gamma _{jj^{\prime }}}$, $\eta _{jj^{\prime
}j^{\prime \prime }}=i\Omega _{jj^{\prime }}\rho _{j^{\prime }j^{\prime
\prime }}$. For simplicity, the Rabi frequency $\Omega _{42}$ ($\Omega _{43}$%
) has been labeled as $\Omega _{24}$ ($\Omega _{34}$) somewhere. The first
term of the right hand of the first equation in Eqns. (\ref{ele2})
disappears when the atomic transition $\left\vert 4\right\rangle
\leftrightarrow \left\vert 2\right\rangle $ is resonantly driven by the
Laser. The second and third terms result from the atom-Laser interactions,
and the forth term is caused by the atom-reservoir coupling. The other two
equations can be interpreted in a similar way. Other density matrix elements
other than the ones given in Eqns. (\ref{ele1}-\ref{ele2}) have no impact on
the ones given in Eqns. (\ref{ele1}-\ref{ele2}) due to the expression of
each term in our master equation.

\section{Heat Currents Controlled by Optical Modulation}

\label{HeatCurrents}

\begin{figure}[t!]
\includegraphics[width=8cm, height=6cm]{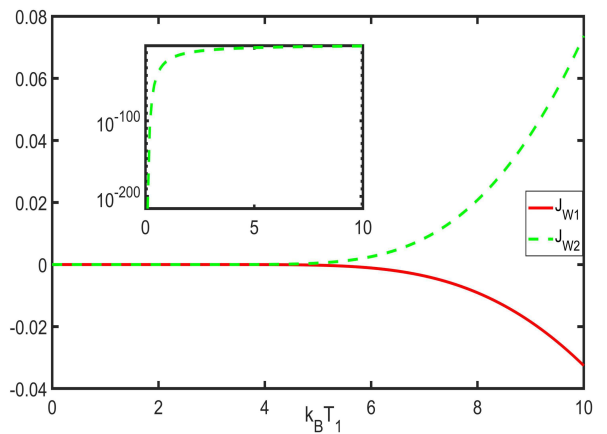}
\caption{Heat currents against the temperature of the reservoir 1 when the
Laser driving the atomic transition $|4\rangle\leftrightarrow|3\rangle$ is
shut off and $T_{2}\rightarrow 0$. The red solid line represents the current
into the reservoir 1 and the green dashed line denotes the current into the
reservoir 2. The parameters are $\protect\omega_{1}$=0, $\protect\omega_2=%
\protect\omega_3=50\protect\gamma_{21}$, $\protect\omega_{4}=70\protect\gamma%
_{21}$, $\Omega_{42}=\protect\gamma_{21}$, $\protect\gamma_{31}=\protect%
\gamma_{21}$, $\protect\omega_{L42}=\protect\omega_{42}$, and $\protect\gamma%
_{42}=\protect\gamma_{43}=0.2\protect\gamma_{21}$.}
\label{current_vs_T1}
\end{figure}

The average energy of the atom is $\left\langle H_{A}\right\rangle
=Tr(H_{A}\rho )$. The time evolution of the average energy going through the
atom is given as%
\begin{equation}
\frac{\partial \left\langle H_{A}\right\rangle }{\partial t}=Tr(H_{A}\dot{%
\rho})=J_{L}^{in}+\sum_{m=1,2}J_{Wm}^{in},
\end{equation}%
with $J_{L}^{in}=-iTr(H_{A}[H_{A-L}^{\prime },\rho ])$, or alternatively $%
J_{L}^{in}=-iTr([H_{A},H_{A-L}^{\prime }]\rho )$, describing the energy flow
from the Laser. Moreover, the energy flow can be divided into two parts as
\begin{eqnarray}
J_{L1}^{in} &=&-iTr(H_{A}[\Omega _{42}\sigma ^{42}+h.c.,\rho ]),  \notag \\
J_{L2}^{in} &=&-iTr(H_{A}[\Omega _{43}\sigma ^{43}+h.c.,\rho ]).
\end{eqnarray}%
The two parts respectively denote the energy flow from either Laser beam.
The heat currents from the reservoirs are%
\begin{eqnarray*}
J_{W1}^{in} &=&Tr(H_{A}\mathcal{L}_{21}[\rho ])+Tr(H_{A}\mathcal{L}%
_{43}[\rho ]), \\
J_{W2}^{in} &=&Tr(H_{A}\mathcal{L}_{31}[\rho ])+Tr(H_{A}\mathcal{L}%
_{42}[\rho ]).
\end{eqnarray*}%
Then the heat current injected into either reservoir is%
\begin{equation*}
J_{Wm}=-J_{Wm}^{in}.
\end{equation*}%
Obviously, $J_{Wm}>0$ implies that the heat is injected into reservoir $m$,
and $J_{Wm}<0$ implies the outflow of heat from the reservoir $m$.
Similarly, the energy flow into the Laser beam is%
\begin{equation*}
J_{Lm}=-J_{Lm}^{in}.
\end{equation*}%
In the steady-state regime, i.e. $\dot{\rho}=0$, the average energy of the
atom does not change against time, and hence $\sum_{m=1,2}(J_{Lm}+J_{Wm})=0$%
. It implies that the steady state of the waveguide-emitter system can be
the nonequilibrium state. In the nonequilibrium state, there are energy
exchanges among the four light fields, i.e. the two guided continuum modes
and the two Laser beams. The conservation of energy is fulfilled.

\begin{figure}[t]
\includegraphics[width=9cm, height=6.5cm]{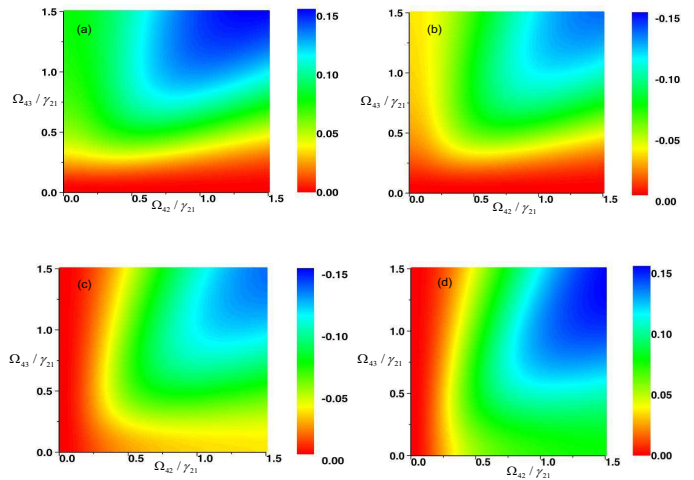}
\caption{Heat currents against the Rabi frequencies of the Laser beams. (a)
and (b) represent the case when $k_{B}T_{1}=1$ and $k_{B}T_{2}=10$. (c) and
(d) represent $k_{B}T_{1}=10$ and $k_{B}T_{2}=1$. (a) and (c) denote the
current $J_{W1}$ against the Rabi frequencies. (b) and (d) denote $J_{W2}$.
The parameters are $\protect\omega _{1}=0$, $\protect\omega _{2}=\protect%
\omega _{3}=50\protect\gamma _{21}$, $\protect\omega _{4}=70\protect\gamma %
_{21}$, $\protect\omega _{L42}=\protect\omega _{42}$, $\protect\omega _{L43}=%
\protect\omega _{43}$, $\protect\gamma _{31}=\protect\gamma _{21}$, and $%
\protect\gamma _{42}=\protect\gamma _{43}=0.2\protect\gamma _{21}$.}
\label{asym_current}
\end{figure}

From the expressions of $J_{Lm}$ and $J_{Wm}$, the density matrix elements
given in Eqns. (\ref{ele1}-\ref{ele2}) are enough to investigate $J_{Lm}$
and $J_{Wm}$. To obtain the steady-state solution, it is necessary to
consider the condition $\sum_{j}\rho _{jj}=1$. This is because the sum of
the four equations in Eqns. (\ref{ele1}) is zero, i.e. $Tr\dot{\rho}=0$.

\subsection{Control of heat conduction by optical modulation}

In our scheme, it is possible to realize the diode-like behavior for the
heat currents, i.e. the asymmetric conduction of the heat against the
temperature gradient, if either of the two Laser beams is shut off. To show
this, we first consider the situation when $\Omega _{43}=0$ and $\Omega
_{42}\neq 0$. In the ideal cases when $T_{1}\rightarrow 0$\ and $T_{2}>0$,
by solving Eqns. (\ref{ele1}-\ref{ele2}), the steady-state density matrix
elements are obtained as
\begin{eqnarray}
\rho _{11} &\rightarrow &\frac{n_{31}+1}{2n_{31}+1},  \notag \\
\rho _{33} &\rightarrow &\frac{n_{31}}{2n_{31}+1},  \label{ideal}
\end{eqnarray}%
and other elements tend to zero. Then, the heat currents into the reservoirs
are obtained as $J_{L1}\rightarrow 0$ and $J_{L2}\rightarrow 0$. Therefore,
the heat conduction from the reservoir 2 to the reservoir 1 is prohibited
although $T_{2}\gg T_{1}$. Conversely, the heat conduction from the
reservoir 1 to the reservoir 2 is allowed if $T_{2}\rightarrow 0$ and $%
T_{1}>0$, which is verified by the numerical simulation as shown in Fig. \ref%
{current_vs_T1}.

The diode-like operation realized when $\Omega _{43}=0$ and $\Omega
_{42}\neq 0$ can be understood by the fact that the heat conduction from one
reservoir to the other is assisted by the cycle of atomic transitions, such
as $\left\vert 1\right\rangle \rightarrow \left\vert 3\right\rangle
\rightarrow \left\vert 4\right\rangle \rightarrow \left\vert 2\right\rangle
\rightarrow \left\vert 1\right\rangle $, $\left\vert 1\right\rangle
\rightarrow \left\vert 2\right\rangle \rightarrow \left\vert 4\right\rangle
\rightarrow \left\vert 3\right\rangle \rightarrow \left\vert 1\right\rangle $%
, or other cycles. The atomic transition from level $\left\vert
3\right\rangle $ to $\left\vert 4\right\rangle $ needs absorbing the energy
from the laser beam or the reservoir 1, and the transition $\left\vert
1\right\rangle $ to $\left\vert 2\right\rangle $ needs absorbing the energy
from the reservoir 1. Obviously, when $T_{1}\rightarrow 0$ and $\Omega
_{43}=0$, both the two atomic transitions will not be realized. Hence, all
the atomic transition cycles that are helpful to the heat conduction between
the two reservoirs can not be achieved. As a result, the heat conduction is
forbidden although $T_{2}\gg T_{1}$. In this case, the system, which can be
considered as a two-level atom (TLA) with levels $\left\vert 3\right\rangle $
and $\left\vert 1\right\rangle $ coupled to the reservoir 2, is in the
thermal equilibrium state. The steady-state solution in Eqns. (\ref{ideal})
implies that the TLA is in the Gibbs thermal state $\frac{e^{-\beta H_{TLA}}%
}{Tr(e^{-\beta H_{TLA}})}$ with $H_{TLA}=\omega _{1}\sigma ^{11}+\omega
_{3}\sigma ^{33}$ representing the energy of the TLA. The effective
temperature of the TLA, i.e. $T_{eff}=\frac{\omega _{31}}{k_{B}}[\ln \frac{%
\rho _{11}}{\rho _{33}}]^{-1}$, tends to $T_{2}$. As $T_{2}$ increases, $%
\rho _{33}$ increases. When $T_{2}\rightarrow \infty $, both $\rho _{11}$
and $\rho _{33}$ tend to $1/2$. For the reversed temperature gradient, i.e. $%
T_{2}\rightarrow 0$ and $T_{1}>0$, the helpful atomic transition cycles are
feasible and hence the heat conduction is realized. It is interesting to
note that the heat conduction absorbed by the cold reservoir is larger than
the conduction left from the hot reservoir in Fig. \ref{current_vs_T1},
which implies the gain of the heat current. This is because the energy of
the Laser beam is converted into the heat during the atomic transition
cycle.
\begin{figure}[t]
\includegraphics[width=8cm, height=10cm]{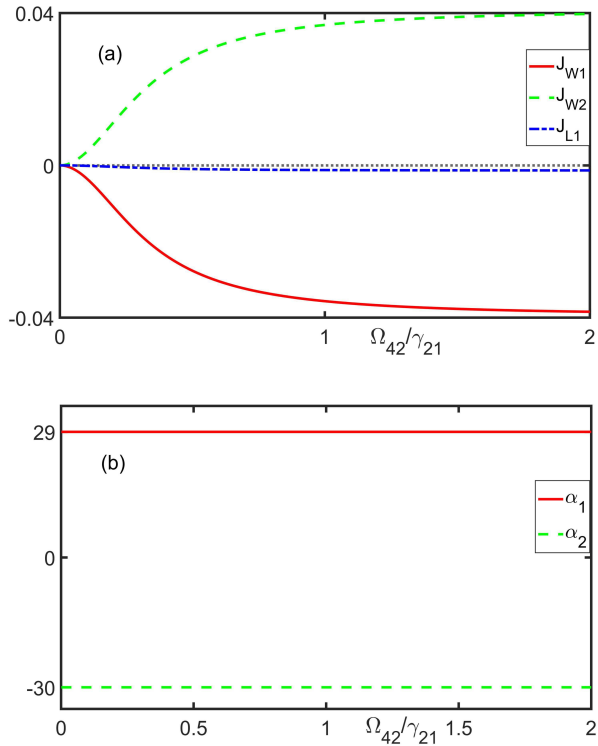}
\caption{Numerical simulation the amplification of the heat currents of the
reservoirs by the Laser. (a) shows $J_{W1}$, $J_{W2}$, and $J_{L1}$ against
the Rabi frequency $\Omega _{42}$, represented by the red solid, green
dashed, and blue dashed dotted lines, respectively. (b) shows the
amplification factors $\protect\alpha _{1}$ and $\protect\alpha _{2}$,
represented by the red solid and green dashed lines, respectively. The
parameters are $k_{B}T_{1}=10$, $k_{B}T_{2}=1$, $\protect\omega _{1}=0$, $%
\protect\omega _{2}=\protect\omega _{3}=60\protect\gamma _{21}$, $\protect%
\omega _{4}=62\protect\gamma _{21}$, $\Omega _{43}=\protect\gamma _{42}=0$, $%
\protect\omega _{L42}=\protect\omega _{42}$, $\protect\gamma _{31}=\protect%
\gamma _{21}$, and $\protect\gamma _{43}=0.1\protect\gamma _{21}$.}
\label{currents_vs_Rabi1}
\end{figure}

Similarly, if the Laser beam driving the atomic transition $\left\vert
4\right\rangle \leftrightarrow \left\vert 3\right\rangle $ is turned on
while the beam driving $\left\vert 4\right\rangle \leftrightarrow \left\vert
2\right\rangle $ is shut off, the heat conduction is forbidden (gained) when
$T_{2}\rightarrow 0$ and $T_{1}>0$ ($T_{1}\rightarrow 0$ and $T_{2}>0$).
Shutting off the different Laser beams forbids the heat currents towards the
different directions, respectively. Therefore, in the ideal case when the
temperature of the cold reservoir approaches zero, the heat conduction
towards either direction can be optimally suppressed while the conduction
along the opposite direction is gained.

Besides the ideal case, when the temperature of the cold reservoir is
nonzero, the pronounced asymmetric conduction of the heat against the
temperature gradient can also be realized. In Fig. \ref{asym_current}, we
plot the heat currents against the Rabi frequencies of the Laser beams. We
consider that the two reservoirs are symmetrical to the four-level atom,
i.e. $\omega _{2}=\omega _{3}$, $\gamma _{21}=\gamma _{31}$, and $\gamma
_{43}=\gamma _{42}$. The plots show the pronounced asymmetric conduction of
the heat when the atom-Laser couplings are pronounced asymmetric, i.e. $\max
\{\Omega _{43},\Omega _{42}\}/\min \{\Omega _{43},\Omega _{42}\}\gg 1$. The
fact that either of the two Laser beams is shut off is the special case of
the asymmetric atom-Laser coupling. The heat conduction from reservoir 1 to
2 is significantly suppressed when $\Omega _{42}\ $is small enough.
Conversely, the conduction from reservoir 2 to 1 is significantly suppressed
when $\Omega _{43}\ $ is small enough. The minimum currents for the
asymmetric conduction into the cold reservoir, i.e. in Fig. \ref%
{asym_current} (a) and (d), are in the order of $10^{-11}$. The heat
transporting towards the allowed directions is\ gained by the Laser beams,
which can be interpreted by comparing the currents of the cold and hot
reservoirs.\ Therefore, the scheme provides the control of the heat
conduction by the optical modulation of the Laser.

\begin{figure}[t!]
\includegraphics[width=8cm, height=5.5cm]{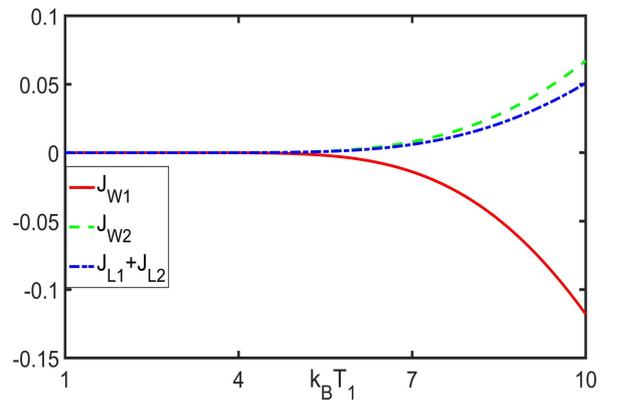}
\caption{The energy flows against the temperature of the reservoir 1 when $%
k_{B}T_{2}=1$. The red solid line represents the heat current into the
reservoir 1, the green dashed line denotes the heat current into the
reservoir 2, and the blue dashed dotted line shows the energy flow into the
Lasers. The parameters are $\protect\omega _{1}$=0, $\protect\omega _{2}=50%
\protect\gamma_{21}$, $\protect\omega _{3}=25\protect\gamma _{21}$, $\protect%
\omega _{4}=60\protect\gamma _{21}$, $\protect\omega _{L43}=\protect\omega %
_{43}$, $\protect\omega _{L42}=\protect\omega _{42}$, $\Omega _{42}=\Omega
_{43}=\protect\gamma _{21}$, $\protect\gamma _{31}=\protect\gamma _{21}$,
and $\protect\gamma _{42}=\protect\gamma _{43}=0.1\protect\gamma _{21}$.}
\label{eigine}
\end{figure}

In an electronic transistor, the currents of the collector and emitter can
be modulated, switched and amplified by the base current. The modulation and
switch of the heat currents by the base current have been realized if we
consider the energy flow of the Laser as the base current. We proceed to
show that the heat currents of the two reservoirs can be amplified by the
Laser, i.e. the heat currents change significantly when the energy flow of
the Laser varies slightly. Followed by the definition in Ref. \cite%
{Quantumthermaltransistor}, the dynamical amplification factor for either
heat current is $\alpha _{m}=\frac{\partial J_{Wm}}{\partial J_{L}}$. The
numerical simulations in Fig. \ref{currents_vs_Rabi1} show that the
significant amplification of the heat currents by the Laser can be realized.
To clearly understand the amplification we assume that the Laser beam
driving the transition $\left\vert 4\right\rangle \leftrightarrow \left\vert
3\right\rangle $ is shut off and the transition $\left\vert 4\right\rangle
\leftrightarrow \left\vert 2\right\rangle $ is decoupled to the reservoir 2
in Fig. \ref{currents_vs_Rabi1}. The decoupling of the waveguide to a
certain atomic transition is feasible for appropriate cutoff frequency of
the waveguide so that the photon with frequency around the corresponding
atomic transition frequency can not transport in the waveguide. In this
case, the heat conduction from reservoir $1$ to $2$ is assisted by the
atomic clockwise transition cycle. The amplification results from the fact
that the frequency of the Laser resonantly driving $\left\vert
4\right\rangle \leftrightarrow \left\vert 2\right\rangle $ is significantly
smaller than the frequencies of the guided photons driving the atom, i.e.
the behaviors of the amplification factors in Fig. \ref{currents_vs_Rabi1}
are $\left\vert \alpha _{1}\right\vert \rightarrow \frac{\omega _{21}-\omega
_{43}}{\omega _{42}}$ and $\left\vert \alpha _{2}\right\vert \rightarrow
\frac{\omega _{31}}{\omega _{42}}$.

\subsection{Engine-like operation}

The goal of a heat engine is to convert the heat energy from the reservoirs
to work. Therefore, the steady-state current $\sum_{m}J_{Wm}<0$, or
alternatively $\sum_{m}J_{Lm}>0$, implies an engine-like operation. For
example, considering the case when the clockwise atomic transition cycle is
dominant in dynamic evolution, it would be feasible to realize a positive
output work when $\omega _{23}>0$. In Fig. \ref{eigine}, we report the
numerical simulation of the engine-like behavior, in which we have taken the
reservoir 1 as the heat reservoir and reservoir 2 as the cold reservoir. It
is easy to understand that the efficiency of the engine $\eta \rightarrow
\frac{\omega _{23}}{\omega _{21}}$ when the atomic transitions $\left\vert
4\right\rangle \leftrightarrow \left\vert 2\right\rangle $ and $\left\vert
4\right\rangle \leftrightarrow \left\vert 3\right\rangle $ are decoupled to
the reservoirs and resonantly driven by the Laser beams. In this case, the
Carnot limit will be realized when $\frac{T_{1}}{T_{2}}=\frac{\omega _{21}}{%
\omega _{31}}$ \cite{maser}.

\section{Conclusions}

\label{conclusions}

In conclusion, we investigate the heat conduction in waveguide QED by the
all-optical modulation. In the weak coupling regime, the master equation of
reduced density operator for the emitter, which is coupled to two
waveguides, is obtained under Born, secular, and Weisskopf-Wigner
approximations. The steady-state heat currents in either waveguide is
derived based on the steady-state solution of the master equation. The
outcomes show that the heat currents can be controlled by the external Laser
beams. The tunable asymmetric conduction of the heat against the temperature
gradient, amplification of heat conduction, engine-like operation are
obtained. In reality, in the waveguide-emitter system, there are inevitable
atomic dissipations to the other modes except for the guided mode. The
dissipation can be incorporated by introducing extra dissipators into the
master equation. We have not investigated this case because the weak
dissipation \cite{pla1,crys} would solely affect the optimal behavior for
the realized thermal operations, which can be easily understood from our
physical interpretation of the realized operations. The thermal management
by optical modulation in waveguide QED paves the way for integrating the
nanoscale thermal devices into a large system.

\section{acknowledgments}

This work is supported by Taishan Scholar Project of Shandong Province
(China) under Grant No.tsqn201812059, the National Natural Science
Foundation of China (11505023, 61675115, 11647171, 11934018,12147146), and
the Strategic Priority Research Program of the Chinese Academy of Sciences
(XDB28000000).

\begin{appendix}

\section*{appendix}

In the appendix part, we obtain the time evolution of $\rho _{A}^{I}(t)$.
The interaction-picture Hamiltoninan has the form of%
\begin{equation*}
H^{I}(t)=H_{A-W}^{I}(t)+H_{A-L}^{I}(t),
\end{equation*}%
with%
\begin{eqnarray*}
&&H_{A-W}^{I}(t) \\
&=&\sum_{k}[g_{21,k}\sigma ^{21}b_{1,k}e^{i(\omega _{21}-\omega
_{1,k})t}+g_{31,k}\sigma ^{31}b_{2,k}e^{i(\omega _{31}-\omega _{2,k})t} \\
&&+g_{43,k}^{\prime }\sigma ^{43}b_{1,k}e^{i(\omega _{43}-\omega
_{1,k})t}+g_{42,k}^{\prime }\sigma ^{42}b_{2,k}e^{i(\omega _{42}-\omega
_{2,k})t}] \\
&&+h.c. \\
&&H_{A-L}^{I}(t) \\
&=&\Omega _{42}\sigma ^{42}e^{i(\omega _{42}-\omega _{L42})t}+\Omega
_{43}\sigma ^{43}e^{i(\omega _{43}-\omega _{L43})t}+h.c.
\end{eqnarray*}%
Under Born approximation, i.e. $\rho _{total}^{I}(t)\simeq \rho
_{A}^{I}(t)\otimes \rho _{W}^{I}(t_{0})$ in the weak-coupling regime, with $%
\rho _{W}^{I}(t_{0})$ the interaction-picture density operator for the
waveguides, the\ Eqn. (\ref{von}) turns out to be

\begin{eqnarray*}
\dot{\rho}_{A}^{I}(t) &=&-i[H_{A-L}^{I}(t),\rho _{A}^{I}(t)] \\
&&-Tr_{W}\int_{t_{0}}^{t}dt^{\prime }[H_{A-L}^{I}(t),[H^{I}(t^{\prime
}),\rho _{A}^{I}(t^{\prime })\otimes \rho _{W}^{I}(t_{0})]].
\end{eqnarray*}%
Then consider the conditions $\left\langle b_{m,k}\right\rangle
=\left\langle b_{m,k}^{\dagger }\right\rangle =0$, $\left\langle
b_{m,k}b_{m^{\prime },k^{\prime }}\right\rangle =\left\langle
b_{m,k}^{\dagger }b_{m^{\prime },k^{\prime }}^{\dagger }\right\rangle =0$, $%
\left\langle b_{m,k}b_{m^{\prime },k^{\prime }}^{\dagger }\right\rangle
=(n_{m,\omega _{k}}+1)\delta _{m,m^{\prime }}\delta _{k,k^{\prime }}$, and $%
\left\langle b_{m,k}^{\dagger }b_{m^{\prime },k^{\prime }}\right\rangle
=n_{m,\omega _{k}}\delta _{m,m^{\prime }}\delta _{k,k^{\prime }}$, we obtain

\begin{widetext}
\begin{eqnarray*}
&&Tr_{W}\int_{t_{0}}^{t}dt^{\prime }[H_{A-L}^{I}(t),[H^{I}(t^{\prime }),\rho
_{A}^{I}(t^{\prime })\otimes \rho _{W}^{I}(t_{0})]]
=\int_{t_{0}}^{t}dt^{\prime }(L_{1}+L_{2})
\end{eqnarray*}

with%
\begin{eqnarray*}
L_{1} &=&\sum_{k}\left\vert g_{21,k}\right\vert ^{2}\left\langle
b_{1,k}b_{1,k}^{\dagger }\right\rangle \{[\sigma ^{21}\sigma ^{12}\rho
_{A}^{I}(t^{\prime })-\sigma ^{12}\rho _{A}^{I}(t^{\prime })\sigma
^{21}]e^{i(\omega _{21}-\omega _{1,k})(t-t^{\prime })}+h.c.\} \\
&&+\sum_{k}\left\vert g_{21,k}\right\vert ^{2}\left\langle b_{1,k}^{\dagger
}b_{1,k}\right\rangle \{[\sigma ^{12}\sigma ^{21}\rho _{A}^{I}(t^{\prime
})-\sigma ^{21}\rho _{A}^{I}(t^{\prime })\sigma ^{12}]e^{-i(\omega
_{21}-\omega _{1,k})(t-t^{\prime })}+h.c.\} \\
&&+\sum_{k}\left\vert g_{43,k}\right\vert ^{2}\left\langle
b_{1,k}b_{1,k}^{\dagger }\right\rangle \{[\sigma ^{43}\sigma ^{34}\rho
_{A}^{I}(t^{\prime })-\sigma ^{34}\rho _{A}^{I}(t^{\prime })\sigma
^{43}]e^{i(\omega _{43}-\omega _{1,k})(t-t^{\prime })}+h.c.\} \\
&&+\sum_{k}\left\vert g_{43,k}\right\vert ^{2}\left\langle b_{1,k}^{\dagger
}b_{1,k}\right\rangle \{[\sigma ^{34}\sigma ^{43}\rho _{A}^{I}(t^{\prime
})-\sigma ^{43}\rho _{A}^{I}(t^{\prime })\sigma ^{34}]e^{-i(\omega
_{43}-\omega _{1,k})(t-t^{\prime })}+h.c.\} \\
&&+\sum_{k}\left\vert g_{31,k}\right\vert ^{2}\left\langle
b_{2,k}b_{2,k}^{\dagger }\right\rangle \{[\sigma ^{31}\sigma ^{13}\rho
_{A}^{I}(t^{\prime })-\sigma ^{13}\rho _{A}^{I}(t^{\prime })\sigma
^{31}]e^{i(\omega _{3}-\omega _{2,k})(t-t^{\prime })}+h.c.\} \\
&&+\sum_{k}\left\vert g_{31,k}\right\vert ^{2}\left\langle b_{2,k}^{\dagger
}b_{2,k}\right\rangle \{[\sigma ^{13}\sigma ^{31}\rho _{A}^{I}(t^{\prime
})-\sigma ^{31}\rho _{A}^{I}(t^{\prime })\sigma ^{13}]e^{-i(\omega
_{3}-\omega _{2,k})(t-t^{\prime })}+h.c.\} \\
&&+\sum_{k}\left\vert g_{42,k}\right\vert ^{2}\left\langle
b_{2,k}b_{2,k}^{\dagger }\right\rangle \{[\sigma ^{42}\sigma ^{24}\rho
_{A}^{I}(t^{\prime })-\sigma ^{24}\rho _{A}^{I}(t^{\prime })\sigma
^{42}]e^{i(\omega _{42}-\omega _{2,k})(t-t^{\prime })}+h.c.\} \\
&&+\sum_{k}\left\vert g_{42,k}\right\vert ^{2}\left\langle b_{2,k}^{\dagger
}b_{2,k}\right\rangle \{[\sigma ^{24}\sigma ^{42}\rho _{A}^{I}(t^{\prime
})-\sigma ^{42}\rho _{A}^{I}(t^{\prime })\sigma ^{24}]e^{-i(\omega
_{42}-\omega _{2,k})(t-t^{\prime })}+h.c.\},
\end{eqnarray*}%
and%
\begin{eqnarray*}
L_{2} &=&\sum_{k}g_{21,k}g_{43,k}^{\ast }e^{i\frac{\omega _{21}-\omega _{43}%
}{2}(t+t^{\prime })}\left\langle b_{1,k}b_{1,k}^{\dagger }\right\rangle
\{[\sigma ^{21}\sigma ^{34}\rho _{A}^{I}(t^{\prime })-\sigma ^{34}\rho
_{A}^{I}(t^{\prime })\sigma ^{21}]e^{i(\frac{\omega _{21}+\omega _{43}}{2}%
-\omega _{1,k})(t-t^{\prime })}+h.c.\} \\
&&+\sum_{k}g_{21,k}g_{43,k}^{\ast }e^{i\frac{\omega _{21}-\omega _{43}}{2}%
(t+t^{\prime })}\left\langle b_{1,k}^{\dagger }b_{1,k}\right\rangle
\{[\sigma ^{34}\sigma ^{21}\rho _{A}^{I}(t^{\prime })-\sigma ^{21}\rho
_{A}^{I}(t^{\prime })\sigma ^{34}]e^{-i(\frac{\omega _{21}+\omega _{43}}{2}%
-\omega _{1,k})(t-t^{\prime })}+h.c.\} \\
&&+\sum_{k}g_{21,k}^{\ast }g_{43,k}e^{-i\frac{\omega _{21}-\omega _{43}}{2}%
(t+t^{\prime })}\left\langle b_{1,k}b_{1,k}^{\dagger }\right\rangle \{\sigma
^{43}\sigma ^{12}\rho _{A}^{I}(t^{\prime })-\sigma ^{12}\rho
_{A}^{I}(t^{\prime })\sigma ^{43}]e^{i(\frac{\omega _{21}+\omega _{43}}{2}%
-\omega _{1,k})(t-t^{\prime })}+h.c.\} \\
&&+\sum_{k}g_{21,k}^{\ast }g_{43,k}e^{-i\frac{\omega _{21}-\omega _{43}}{2}%
(t+t^{\prime })}\left\langle b_{1,k}^{\dagger }b_{1,k}\right\rangle
\{[\sigma ^{12}\sigma ^{43}\rho _{A}^{I}(t^{\prime })-\sigma ^{43}\rho
_{A}^{I}(t^{\prime })\sigma ^{12}]e^{-i(\frac{\omega _{21}+\omega _{43}}{2}%
-\omega _{1,k})(t-t^{\prime })}+h.c.\} \\
&&+\sum_{k}g_{31,k}g_{42,k}^{\ast }e^{i\frac{\omega _{31}-\omega _{42}}{2}%
(t+t^{\prime })}\left\langle b_{2,k}b_{2,k}^{\dagger }\right\rangle
\{[\sigma ^{31}\sigma ^{24}\rho _{A}^{I}(t^{\prime })-\sigma ^{24}\rho
_{A}^{I}(t^{\prime })\sigma ^{31}]e^{i(\frac{\omega _{31}+\omega _{42}}{2}%
-\omega _{2,k})(t-t^{\prime })}+h.c.\} \\
&&+\sum_{k}g_{31,k}g_{42,k}^{\ast }e^{i\frac{\omega _{31}-\omega _{42}}{2}%
(t+t^{\prime })}\left\langle b_{2,k}^{\dagger }b_{2,k}\right\rangle
\{[\sigma ^{24}\sigma ^{31}\rho _{A}^{I}(t^{\prime })-\sigma ^{31}\rho
_{A}^{I}(t^{\prime })\sigma ^{24}]e^{-i(\frac{\omega _{31}+\omega _{42}}{2}%
-\omega _{2,k})(t-t^{\prime })}]+h.c.\} \\
&&+\sum_{k}g_{31,k}^{\ast }g_{42,k}e^{-i\frac{\omega _{31}-\omega _{42}}{2}%
(t+t^{\prime })}\left\langle b_{2,k}b_{2,k}^{\dagger }\right\rangle
\{[\sigma ^{42}\sigma ^{13}\rho _{A}^{I}(t^{\prime })-\sigma ^{13}\rho
_{A}^{I}(t^{\prime })\sigma ^{42}]e^{i(\frac{\omega _{31}+\omega _{42}}{2}%
-\omega _{2,k})(t-t^{\prime })}+h.c.\} \\
&&+\sum_{k}g_{31,k}^{\ast }g_{42,k}e^{-i\frac{\omega _{31}-\omega _{42}}{2}%
(t+t^{\prime })}\left\langle b_{2,k}^{\dagger }b_{2,k}\right\rangle
\{[\sigma ^{13}\sigma ^{42}\rho _{A}^{I}(t^{\prime })-\sigma ^{42}\rho
_{A}^{I}(t^{\prime })\sigma ^{13}]e^{-i(\frac{\omega _{31}+\omega _{42}}{2}%
-\omega _{2,k})(t-t^{\prime })}+h.c.\}.
\end{eqnarray*}
\end{widetext}The first four lines of $L_{2}$ imply the indirect
interactions between atomic transitions $\left\vert 2\right\rangle
\leftrightarrow \left\vert 1\right\rangle $ and $\left\vert 4\right\rangle
\leftrightarrow \left\vert 3\right\rangle $ due to the fact that both the
two transitions are driven by the intermediate waveguide 1. Similarly, the
last four lines of $L_{2}$ imply the indirect interactions between atomic
transitions $\left\vert 3\right\rangle \leftrightarrow \left\vert
1\right\rangle $ and $\left\vert 4\right\rangle \leftrightarrow \left\vert
2\right\rangle $ intermediated by waveguide 2. However, the part of $L_{2}$
will be neglected based on two facts. The terms containing $\sigma
^{mn}\sigma ^{m^{\prime }n\prime }$ with $n\neq m^{\prime }$ can be removed
directly. The other terms will be neglected under the secular approximation
in the following step because we have assumed that the values of elements in
the set $\{\omega _{43},\omega _{42}\}$ are obviously different from the
ones in $\{\omega _{21},\omega _{31}\}$. For example, after the following
integration under Weisskopf-Wigner approximation, the first line of $L_{2}$
is multiplied by $e^{i(\omega _{21}-\omega _{43})t}$ and hence it is
neglected.

The sum of the discrete wavevector $k$ can be treated as $\lim_{\Delta
k\rightarrow 0}\sum_{k}f(k)\Delta k\rightarrow \int_{0}^{+\infty }f(k)dk$,
with $\Delta k=\frac{2\pi }{L}$ under the periodic boundary condition for a
confined 1D space with length $L$. The lower limit of the integral is $0$
rather than $-\infty $ because one can consider that the photon propagates
unidirectionally in the side-coupled waveguide, see Ref. \cite{fanthero} for details. For
the emission spectrum due to the atom-waveguide interaction, the intensity
of the emitted light is centered about the corresponding atomic transition
frequencies. Thus, the term $\left\vert g_{mn,k}\right\vert ^{2}\left\langle
b_{1,k}b_{1,k}^{\dagger }\right\rangle $ ($\left\vert g_{mn,k}\right\vert
^{2}\left\langle b_{1,k}^{\dagger }b_{1,k}\right\rangle $)\ in $L_{1}$ is
approximately $\left\vert g_{mn,\omega _{mn}}\right\vert ^{2}(n_{\omega
_{mn}}+1)$ ($\left\vert g_{mn,\omega _{mn}}\right\vert ^{2}n_{\omega _{mn}}$%
), and the term $g_{mn,k}g_{lh,k}^{\ast }\left\langle b_{1,k}b_{1,k}^{\dagger
}\right\rangle $ ($g_{mn,k}g_{lh,k}^{\ast }\left\langle b_{1,k}^{\dagger
}b_{1,k}\right\rangle $)\ in $L_{2}$ is approximately $g_{mn,\omega
_{mn}}g_{lh,\omega _{lh}}^{\ast }(n_{\frac{\omega _{mn}+\omega _{lh}}{2}}+1)$
($g_{mn,\omega _{mn}}g_{lh,\omega _{lh}}^{\ast }n_{\frac{\omega _{mn}+\omega
_{lh}}{2}}$). Consider the linear dispersion relationship of the guided
photon, the integral $\int_{0}^{\infty }dke^{i(\zeta -\omega
_{m,k})(t-t^{\prime })}\rightarrow \frac{1}{v_{g}}\int_{-\infty }^{\infty
}d(\omega _{m,k}-\zeta )e^{i(\omega _{m,k}-\zeta )(t^{\prime }-t)}=\frac{%
2\pi }{v_{g}}\delta (t^{\prime }-t)$, with $v_{g}$ the photonic group
velocity and considered unit here. Therefore, under Weisskopf-Wigner and
secular approximations, the time evolution of interaction-picture reduced
density operator for the atom reduces to
\begin{widetext}
\begin{eqnarray*}
\dot{\rho}_{A}^{I}(t) &=&-i[H_{A-L}^{I}(t),\rho _{A}^{I}(t)] \\
&&+\gamma _{21}(n_{\omega _{21},1}+1)(\sigma ^{12}\rho _{A}^{I}(t)\sigma
^{21}-\frac{1}{2}\{\sigma ^{22},\rho _{A}^{I}(t)\})+\gamma _{21}n_{\omega _{21},1}(\sigma ^{21}\rho _{A}^{I}(t)\sigma ^{12}-\frac{1}{2}%
\{\sigma ^{11},\rho _{A}^{I}(t)\}) \\
&&+\gamma _{43}(n_{\omega _{43},1}+1)(\sigma ^{34}\rho _{A}^{I}(t)\sigma
^{43}-\frac{1}{2}\{\sigma ^{44},\rho _{A}^{I}(t)\})+\gamma _{43}n_{\omega _{43},1}(\sigma ^{43}\rho _{A}^{I}(t)\sigma ^{34}-%
\frac{1}{2}\{\sigma ^{33},\rho _{A}^{I}(t)\}) \\
&&+\gamma _{31}(n_{\omega _{31},2}+1)(\sigma ^{13}\rho _{A}^{I}(t)\sigma
^{31}-\frac{1}{2}\{\sigma ^{33},\rho _{A}^{I}(t)\})+\gamma _{31}n_{\omega _{31},2}(\sigma ^{31}\rho _{A}^{I}(t)\sigma ^{13}-%
\frac{1}{2}\{\sigma ^{11},\rho _{A}^{I}(t)\}) \\
&&+\gamma _{42}(n_{\omega _{42},2}+1)(\sigma ^{24}\rho _{A}^{I}(t)\sigma
^{42}-\frac{1}{2}\{\sigma ^{44},\rho _{A}^{I}(t)\})+\gamma _{42}n_{\omega _{42},2}(\sigma ^{42}\rho _{A}^{I}(t)\sigma ^{24}-%
\frac{1}{2}\{\sigma ^{22},\rho _{A}^{I}(t)\}),
\end{eqnarray*}%
\end{widetext}
where $\gamma _{mn}=\left\vert g_{mn,\omega _{mn}}\sqrt{L}\right\vert ^{2}$.
The factor $\sqrt{L}$ is interpreted as the normalization constant arising
from the conversion of the sum into the integral, with the details shown in
Ref. \cite{faninout}.

\end{appendix}

\end{document}